\documentclass[nofootinbib,notitlepage,superscriptaddress,10pt,aps,pra]{revtex4-1}

\usepackage[utf8x]{inputenc}
\usepackage{amsfonts}
\usepackage{amsmath}
\usepackage{amssymb}
\usepackage{graphicx}
\usepackage[normalem]{ulem}
\usepackage{epsf,epsfig}
\usepackage{psfrag}
\usepackage{color}
\usepackage{multirow}
\usepackage{diagbox}
\usepackage{dsfont}
\usepackage{cancel}

\makeatletter
\newcommand\stroke[1]{\mathpalette\stroke@aux{#1}}
\def\stroke@aux#1#2{%
 \ooalign{%
  \hfil$#1^{\;\, \_\hspace{-0.05cm}\_}$\hfil\cr
  \hfil$#1#2$\hfil\crp
 }%
}
\makeatother

\begin{document}

\title{Multi-particle systems in quantum spacetime\\ and a novel challenge for center-of-mass motion}

\author{Giovanni Amelino-Camelia}
\affiliation{Dipartimento di Fisica Ettore Pancini, Università di Napoli ”Federico II”, and INFN, Sezione di Napoli, Complesso Univ. Monte S. Angelo, I-80126 Napoli, Italy}

\author{Valerio Astuti}
\affiliation{Dipartimento di Fisica, Universit\`a di Roma ``La Sapienza", P.le A. Moro 2, 00185 Roma, Italy}
\affiliation{INFN, Sez.~Roma1, P.le A. Moro 2, 00185 Roma, Italy}

\author{Michelangelo Palmisano}
\affiliation{Dipartimento di Fisica, Universit\`a di Roma ``La Sapienza", P.le A. Moro 2, 00185 Roma, Italy}
\affiliation{INFN, Sez.~Roma1, P.le A. Moro 2, 00185 Roma, Italy}

\author{Michele Ronco}
\affiliation{Laboratoire de Physiqe Nucléaire et de Hautes Energies (LPNHE) UPMC, Case courrier 200, 4 place Jussieu, F-75005 Paris, France}
\affiliation{UPMC, Case courrier 200, 4 place Jussieu, F-75005 Paris, France}

\begin{abstract}
In recent times there has been considerable interest in scenarios for quantum gravity in which
particle kinematics is affected nonlinearly by the Planck scale, with encouraging results
for the phenomenological prospects, but also some concerns that the nonlinearities might produce
pathological properties for composite/multiparticle systems. We here focus on kinematics
in the $\kappa$-Minkowski noncommutative spacetime, the quantum spacetime which has
been most studied from this perspective, and compare the implications of the alternative
descriptions of the total momentum of a multiparticle system which have been so far proposed.
We provide evidence suggesting that priority should be given to defining the total momentum as the
standard linear sum of the momenta
of the particles composing the system. We also uncover a previously unnoticed feature concerning
some (minute but conceptually important) effects on center-of-mass motion due to properties of the motion of the constituents relative to the center of mass.
\end{abstract}

\maketitle

\begin{flushleft}

\section{Introduction}
Various semi-heuristic arguments \cite{BergSmit1982, DFR1995, amelino2002IJMPD} suggest that
the Planck scale should affect relativistic kinematics by introducing some nonlinearities. This is also found deductively
in some of the formalisms under consideration in quantum-gravity research, such as the $\kappa$-Minkowski noncommutative
spacetime \cite{MajRue1994, GGV2013} and some models formulated within the ``relative-locality framework" \cite{reloc2011, reloc2013}.
Also noteworthy is the fact that in 2+1D quantum gravity the Planck scale does indeed introduce some nonlinearities
 in relativistic kinematics \cite{freidelElivinePRL}.
Further reasons of interest in this possibility come from the phenomenology side: while most other conjectured Planck-scale
effects are expected to remain untestable for the foreseeable future, it happens to be the case
that nonlinear deformations of relativistic kinematics, even when introduced with Planck-scale suppression,
are within the reach of some present and forthcoming experiments and observations \cite{gacLivingReview}.
However, due to the complexity of the relevant formalisms, it is not fully established whether models with
nonlinear deformations of relativistic kinematics truly are a viable possibility for consistent physics:
for systems of only a small number of particles the phenomenological consequences are plausible and interestingly within reach,
but for systems composed of a large number of particles there are legitimate concerns \cite{Soccer 1, Soccer 2, Soccer 3} that the nonlinearities,
although minute for each individual particle composing the system,
could somehow add up to levels that might be in clear conflict with established experimental facts.

We here explore some of these issues through an analysis of multi-particle systems in the $\kappa$-Minkowski spacetime, a 3+1-dimensional spacetime whose key property is the non-commutativity of its coordinates:
\begin{equation}
\left[x_i,x_0 \right]=i\ell x_i ~,~~~
\left[x_i,x_j \right]=0,
\end{equation}
where $x_0$ is the time, $x_i$ are the three spatial coordinates and $\ell$ is a deformation parameter usually assumed to be of the order of  the Planck length.
A large literature has been devoted to the fact that the symmetries of this quantum space-time are described by the $\kappa$-Poincaré Hopf Algebra which is a non-linear deformation of the standard Poincaré Lie Algebra. Of potential relevance for the description of multiparticle systems in $\kappa$-Minkowski is the observation that, as reviewed in the next section, the structure of the $\kappa$-Poincaré algebra
provides room for a nonlinear deformation of the law of composition of momenta. Several studies (see, {\it e.g.} \cite{gacLivingReview, nonlin1, Soccer 1, Soccer 3})
have compared descriptions of total momentum in $\kappa$-Minkowski based on such nonlinear laws of composition to the standard one based on the linear composition of momenta. The main element of novelty we here contribute to this debate concerns the needed consistency between
the notion of total momentum and the notion of center-of-mass position coordinates.
Some valuable insight is gained when this observation is combined
with the requirement that the description of macroscopic systems (systems with a very large number of composing particles)
should not be pathological: nonlinearities in the definition of total momentum are in general admissible, but only if they do not grow pathologically for macroscopic systems.
Our findings suggest that priority should be given to defining the total momentum as the
standard linear sum of the momenta
of the particles composing the system.

Our investigations also uncover a feature which had previously gone unnoticed:
in $\kappa$-Minkowski
the relative motion and the center of mass motion do not fully decouple. While this feature appears likely
 to produce only minute (untestably small) effects, it could be rather significant conceptually and
 it does not appear to be a peculiarity of $\kappa$-Minkowski: it may well be present in other quantum spacetimes, in which
 however techniques suitable for exposing it have not yet been developed.

 In the next section we briefly review some relevant properties of the $\kappa$-Poincar\'e Hopf algebra.
In section \ref{sec3} we show how the nonlinear $\kappa$-Poincaré-inspired definition of total momentum leads to
a pathological description of macroscopic bodies. In Section \ref{sec4} we propose a new symmetry algebra for the composite system, and expose some advantages of the standard linear definition of total momentum. Some closing
 remarks are offered in Section \ref{sec5}.

\section{The $k$-Poincaré algebra}
\label{sec2}

We shall adopt the most common description of $\kappa$-Poincaré Hopf algebra which is based on the so called bicross product basis,
 such that the commutation relations among algebra generators take the following form
\begin{alignat}{2}\begin{aligned}\label{bicross}
&[P_\mu, P_\nu ]= 0, \quad &&[ R_i, P_0 ]= 0,\\
&[ R_i,P_j ]= i\epsilon_{ijk}P_k, \quad &&[ R_i, R_j ]= i\epsilon_{ijk} R_k, \\
&[ R_i, N_j ]= -i\epsilon_{ijk}  N_k, \quad &&[N_{i},P_{0}]=iP_{i}, \\
&[N_{i},P_{j}] = i\delta_{ij}\Bigl(\frac{1-e^{-2\ell P_{0}}}{2\ell}+\frac{\ell}{2}{| \vec{P} | }^{2}\Bigr)-i\ell P_{i}P_{j}, \quad &&[N_{i},N_{j}]= -i\epsilon_{ijk}R_{k},
\end{aligned}\end{alignat}
where $R_i$ are rotation generators, $N_i$ are boost generators, and $P_{\mu}$ are translation generators.
These also imply a deformaton of the mass Casimir
\begin{equation}
\label{deformedcasimir}
m^2 = \Bigl(\frac{2}{\ell} \sinh{ \frac{  \ell  }{  2 }  }P_0\Bigr)^2 - e^{\ell P_0} |\vec{P}|^2 \, ,
\end{equation}
which indeed is an invariant of the symmetry algebra \eqref{bicross}.

The fact that in the standard Poincarè Lie algebra the action of generators on products of functions
is governed by the  Leibniz rule ($X\triangleright (fg)=(X\triangleright f)g+f( X\triangleright g)$) can be expressed in the
language of Hopf algebras by stating that the coproducts are primitive:
\begin{align}
\begin{split}
&\Delta P_0=P_0\otimes 1 +1 \otimes P_0\\
&\Delta P_i=P_i\otimes 1 +1 \otimes P_i\\
&\Delta N_i=N_i\otimes 1+1 \otimes N_i\\
&\Delta R_i=R_i \otimes 1+1\otimes R_i.
\end{split}
\end{align}

It is well established \cite{MajRue1994} that for the generators of the $\kappa$-Poincaré Hopf algebra one must adopt deformations of the Leibniz rule, those codified in the following non-primitive coproducts:
\begin{align}
\begin{split}
\label{deformedcoproducts}
&\Delta P_0=P_0\otimes 1 +1 \otimes P_0\\
&\Delta P_i=P_i\otimes 1 +e^{-\ell P_0} \otimes P_i\\
&\Delta N_i=N_i\otimes 1+e^{-\ell P_0} \otimes N_i+\ell \epsilon_{ijk} P_j\otimes R_k\\
&\Delta R_i=R_i \otimes 1+1\otimes R_i.
\end{split}
\end{align}
Such that, for example, $P_i\triangleright \left(f(x) g(x)  \right) = \left( P_i \triangleright f(x)\right) g(x) + \left( e^{-\ell P_0} \triangleright f(x) \right)  \left( P_i \triangleright g(x) \right) $.

The coproducts of translation generators indeed suggest
 a nonlinear
law of composition of momenta, which for example, in the case of two particles with momenta $p$ and $k$ gives
\begin{equation}
\label{deformedconservation}
(p \oplus k)_0 = p_{0}+k_{0} \qquad (p \oplus k)_i = p_{i}+e^{- \ell p_{0}}k_{i} \, .
\end{equation}
 However, while for microscopic particles $p_{0} \ll 1/\ell$
(if $\ell$ is indeed of the order of the Planck length), if the two momenta being composed are momenta of macroscopic bodies
this composition law is evidently pathological. This shall be a key observation for
our description of multiparticle systems.

Also relevant for our investigations is the well-established fact that one can describe the
symmetries of $\kappa$-Minkowski in terms of the $\kappa$-Poincaré
Hopf algebra \cite{MajRue1994}. For example, from the following rules of action
of the translation generators on coordinates:
$$P_0 \triangleright x_0 = i  ,\quad  P_i \triangleright x_j = -i \delta_{ij}.$$
one finds that the $\kappa$-Minkowski commutation relations are covariant:
$$P_i\triangleright [x_j,x_0]=P_i\triangleright x_j x_0 - P_i\triangleright x_0 x_j =$$
$$=-i \delta_{ij} x_0-(e^{-\ell P_0}\triangleright x_0)P_i\triangleright x_j=-i \delta_{ij} x_0+ i \delta_{ij} (1-\ell P_0)\triangleright x_0$$
$$=\ell \delta_{ij} =i \ell P_i\triangleright x_j$$
In what here follows important roles are played by the
associated deformed Heisenberg algebra \cite{HeisAlg},
$$[P_0,x_0]=i, \qquad [P_0,x_j]=0,$$
$$[P_i,x_0]=-i\ell P_i, \qquad [P_i,x_j]=-i \delta_{ij} ~,$$
and by the representation of boost generators $N_i$,
\begin{equation}
\label{diffreprboost}
N_i = x_i \Bigl({1-e^{-2\ell P_0} \over 2\ell}+{\ell \over 2}| \vec{P} |^2\Bigr)-x_0 P_i.
\end{equation}

\section{Troublesome deformation of center of mass coordinates from deformation of momentum-composition law}
\label{sec3}

Previous studies have focused on the fact that descriptions of total momentum based on the
$k$-Poincaré-deformed composition law could produce a paradoxical description of the total momentum
of macroscopic bodies. We here intend to notice that descriptions of total momentum based on the
$k$-Poincaré-deformed composition law also lead to an unsatisfactory description of the center-of-mass coordinates,
an issue which will play a key role in our following observations.

For simplicity we work in a 1+1-dimensional $k$-Minkowski and
focus on only two identical particles $A$ and $B$, describing their total momentum in
terms of the $k$-Poincaré-deformed composition law:
\begin{align}
\begin{split}
\label{generatoritotalik}
&P_0^T=P_0^A+P_0^B,\\
&P_1^T=P_1^A+e^{-\ell P_0^A}P_1^B.
\end{split}
\end{align}
Our requirement is that there should be a pair of coordinates, $x_0^T$, $x_1^T$, describing the position of the center of mass
of the system. We expect to have that $P_\mu^T$ and $x_\mu^T$ are dual, {\it i.e.} they close a (possibly deformed) Heisenberg Algebra.

The standard choice of center-of-mass coordinates:
\begin{align}
\begin{split}
&x_0^T={x_0^A+x_0^B \over 2},\\
&x_1^T={x_1^A+x_1^B\over 2}
\end{split}
\end{align}
is clearly not suitable for (\ref{generatoritotalik}), since one has that:
\begin{align}
\begin{split}
[P_1^T,x_1^T]&=[P_1^A+e^{-\ell P_0^A}P_1^B,{x_1^A+x_1^B\over 2}]\\&=-{i \over 2}(1+e^{-\ell P_0^A})
\end{split}
\end{align}
and therefore there is no closed (however deformed) center-of-mass Heisenberg algebra.

It is easy to show that a modified definition of the center-of-mass coordinates governed by a closed (deformed) center-of-mass Heisenberg algebra can be found in the form:
\begin{align}
\begin{split}
&x_0^T={x_0^A+x_0^B+\frac{\ell}{2} \left(x_1^A P_1^A + P_1^A x_1^A  \right) \over 2},\\
&x_1^T={x_1^A+e^{\ell P_0^A}x_1^B \over 2}
\end{split}
\end{align}
The potential relevance of these coordinates could also be suggested by the fact that
they are still $k$-Minkowski coordinates, like those of the constituent particles:
\begin{align}
\begin{split}
[x_1^T,x_0^T]&={1 \over 4}[x_1^A,x_0^A]+{1 \over 4}[x_1^A,\ell x_1^A P_1^A]+{1 \over 4}[e^{\ell P_0^A}x_1^B,x_0^A]+{1 \over 4}[e^{\ell P_0^A}x_1^B,x_0^B]\\&=i{\ell \over 4}x_1^A+i{\ell \over 4}x_1^A+i{\ell \over 4}e^{\ell P_0^A}x_1^B+i{\ell \over 4}e^{\ell P_0^A}x_1^B\\&=i\ell\Bigl({x_1^A+e^{\ell P_0^A}x_1^B \over 2}\Bigr)\\&=i\ell x_1^T
\end{split}
\end{align}
These $\{x_1^T,x_0^T\}$ together with $\{P_1^T,P_0^T\}$ also satisfy the $k$-deformed Heisenberg algebra:
\begin{alignat}{2}
\begin{aligned}
&[P_0^T,x_0^T]=i,\qquad && [P_0^T,x_1^T]=0,\\
&[P_1^T,x_0^T]=-i\ell P_1^T, \qquad && [P_1^T,x_1^T]=-i,
\end{aligned}
\end{alignat}

While all these might be technically reassuring, the emerging physical picture is not encouraging:
if these were two macroscopic bodies one would get an unsatisfactory picture not only for
the total momentum specified by (\ref{generatoritotalik}) but
also for the spatial coordinates of the center of mass: if $P_0^A \gg 1/\ell$ (not uncommon for a macroscopic body
if $\ell$ is of the order of the Planck length) then $x_1^T={x_1^A+e^{\ell P_0^A}x_1^B \over 2}$ would assign
a $x_1$ coordinate to the center of mass which either is very close to $x_1^A$, ignoring the  $x_1^B$ contribution
completely (if $\ell < 0$),
or is close to being proportional to $x_1^B$, ignoring the  $x_1^A$ contribution
completely (if $\ell > 0$).

\section{Symmetry algebra of a system of 2 identical particles}
\label{sec4}
One our way to an alternative picture,
let us then reconsider two particles $A$ and $B$ of equal mass $m$ and coordinates $x_\mu^A$ and $x_\mu^B$ satisfying
($\mu \in \{ 0,1 \}$)
\begin{equation}
[x_1^{A,B},x_0^{A,B}]=i\ell x_1^{A,B}, \qquad [x_{\mu}^A,x_{\nu}^B]=0.
\end{equation}
For both of them we can build a $\kappa$-Poincaré algebra with generators $\{P_\mu^{A,B},N^{A,B}\}$, where in particular the translation generators $P_\mu^{A,B}$ act on the respective sets of plane waves as
\begin{equation}
P_\mu^{A,B}\triangleright e^{ik_1^{A,B} x_1^{A,B}}e^{-ik_0^{A,B} x_0^{A,B}}=k_\mu^{A,B}e^{ik_1^{A,B} x_1^{A,B}}e^{-ik_0^{A,B} x_0^{A,B}}.
\end{equation}
In other words the plane waves $e^{ik_1^A x_1^A}e^{-ik_0^A x_0^A}$ and $e^{ik_1^B x_1^B}e^{-ik_0^B x_0^B}$ diagonalize the action of the operators $P_\mu^A$ and $P_\mu^B$, respectively.
As in the standard relativistic case, we would like to describe our two-particle system by separating it in a ``total" part and a ``relative" part.
To do so let us recall the undeformed center-of-mass coordinates
\begin{equation}
x_\mu^T={x_\mu^A+x_\mu^B \over 2}
\end{equation}
and the relative ones
\begin{equation}
x_\mu^R={x_\mu^A-x_\mu^B \over 2} \, .
\end{equation}
It is evidently noteworthy that these center-of-mass coordinates satisfy a $2\kappa$-Minkowski commutation relation
\begin{equation}
[x_1^T,x_0^T]={1 \over 4}([x_1^A,x_0^A]+[x_1^B,x_0^B])=i{\ell \over 2}x_1^T \, .
\end{equation}
Unlike the case of the center-of-mass coordinates conjugated with the $\kappa$-deformed total momentum, these
center-of-mass coordinates give rise to a truly reassuring picture, in which our composite system
composed of two particles is governed by a halved deformation parameter.
In this way as the number of particles composing the system grows, the deformation effects should become weaker and weaker, thus recovering the standard Poincaré/Galilei physics on macroscopic scales.
The only challenges  reside
in the interface between center-of-mass degrees of freedom and degreees of freedom of the motion relative
to the center of mass, as one can start to appreciate by noticing that
\begin{align}
\begin{split}
&[x_1^T,x_0^T]=i{\ell \over 2}x_1^T, \qquad [x_1^R,x_0^T]=i{\ell \over 2}x_1^R,\\&
[x_1^T,x_0^R]=i{\ell \over 2} x_1^R, \qquad [x_1^R, x_0^R]=i{\ell \over 2} x_1^T. \end{split}
\end{align}
This is further exposed by studying the generators for
center-of-mass and relative-motion symmetries which can be paired with these coordinates. En route to those
we consider the product of two plane waves, one of ''type $A$" and one of ''type $B$":
\begin{equation}\label{abord}
e^{ik_1^A x_1^A}e^{-ik_0^A x_0^A}e^{ik_1^B x_1^B}e^{-ik_0^B x_0^B}
\end{equation}
which we rewrite using the relationship between the coordinates $x_\mu^{A,B}$ and the coordinates $x_\mu^{T,R}$,
finding
\begin{equation}\label{trord}
e^{ik_1^T x_1^T}e^{ik_1^R x_1^R}e^{-ik_0^T x_0^T}e^{-ik_0^R x_0^R}
\end{equation}
with
\begin{equation}
k_\mu^{T}=k_\mu^A+k_\mu^B, \qquad k_\mu^R=k_\mu^A-k_\mu^B.
\end{equation}
It is therefore natural to adopt the following intuitive (standard) definition of translation generators
\begin{equation}
\label{lineartranslation}
P_\mu^T=P_\mu^A+P_\mu^B, \qquad P_\mu^R=P_\mu^A-P_\mu^B
\end{equation}
for which one interestingly finds that
\begin{align}
&P_\mu^{T,R}\triangleright e^{ik_1^T x_1^T}e^{ik_1^R x_1^R}e^{-ik_0^T x_0^T}e^{-ik_0^R x_0^R}=k_\mu^{T,R}e^{ik_1^T x_1^T}e^{ik_1^R x_1^R}e^{-ik_0^T x_0^T}e^{-ik_0^R x_0^R},
\end{align}
where we took into account the canonical action of our operators ($P_{0} \triangleright x_{0} = i$, $P_{1} \triangleright x_{1} = -i$).

It is then straightforward to verify that
\begin{alignat}{2}
\begin{aligned}
\label{heisenberglinear}
&[P_0^T,x_0^T]=i, \qquad &&[P_0^T,x_0^R]=0,\\
&[P_0^T,x_1^T]=0,\qquad &&[P_0^T,x_1^R]=0,\\
&[P_0^R,x_0^T]=0,\qquad &&[P_0^R,x_0^R]=i,\\
&[P_0^R,x_1^T]=0,\qquad &&[P_0^R,x_1^R]=0,\\
&[P_1^T,x_0^T]=-i{\ell \over 2}P_1^T,\qquad &&[P_1^T,x_0^R]=-i{\ell \over 2}P_1^R,\\
&[P_1^T,x_1^T]=-i,\qquad &&[P_1^T,x_1^R]=0\\
&[P_1^R,x_0^T]=-i{\ell \over 2}P_1^R,\qquad &&[P_1^R,x_0^R]=-i{\ell \over 2}P_1^T,\\
&[P_1^R,x_1^T]=0,\qquad &&[P_1^R,x_1^R]=-i
\end{aligned}
\end{alignat}
and derive the following coproducts
\begin{align}
\begin{split}
\label{coprodottilinear}
&\Delta(P_0^{T,R})=P_0^{T,R}\otimes 1+1\otimes P_0^{T,R},\\
&\Delta(P_1^T)=P_1^T\otimes1+e^{-{\ell \over 2}P_0^T}\cosh {\ell \over 2}P_0^R \otimes P_1^T-e^{-{\ell \over 2}P_0^T}\sinh {\ell \over 2}P_0^R \otimes P_1^R,\\
&\Delta(P_1^R)=P_1^R\otimes1+e^{-{\ell \over 2}P_0^T}\cosh {\ell \over 2}P_0^R \otimes P_1^R-e^{-{\ell \over 2}P_0^T}\sinh {\ell \over 2}P_0^R \otimes P_1^T.
\end{split}
\end{align}
Notice how the first two coproducts (as well as the relevant commutators in the deformed Heisenberg algebra \eqref{heisenberglinear}) produce a picture in which the motion relative to the center of mass does not decouple
from the motion of the center of mass itself; however, this conceptually important feature might have little or no observable
consequences, since for small energy-momentum of the relative motion the decoupling is restored and for typical
macroscopic bodies one does expect small energy-momentum of the relative motion.
We find that, when one can neglect the energy-momentum of the relative motion,
the first two coproducts (as well as the relevant commutators in the deformed Heisenberg algebra \eqref{heisenberglinear}) reduce to $2\kappa$-Poincaré ones.

For what concerns boost generators, we get a promising picture by simply emulating the definition of our total and relative translation generators, {\it i.e.}:
\begin{align}
\begin{split}
\label{NTNRlinear}
&N^T=N^A+N^B,\\
&N^R=N^A-N^B.
\end{split}
\end{align}

Using the single particle boost generators representation \eqref{diffreprboost}, the equations in \eqref{NTNRlinear} can be explicitly written as
\begin{align}
\begin{split}
N^T&=x_1^T\Bigl[{1-e^{-\ell P_0^T}\cosh \ell P_0^R \over \ell}+{\ell \over 4}(P_1^{T^2}+P_1^{R^2})\Bigr]+x_1^R\Bigl({e^{-\ell P_0^T}\sinh \ell P_0^R\over \ell}+{\ell\over 2}P_1^TP_1^R\Bigr)-x_0^TP_1^T-x_0^RP_1^R.
\end{split}
\end{align}
and
\begin{align}
\begin{split}
N^R&=x_1^T\Bigl({e^{-\ell P_0^T}\sinh \ell P_0^R\over \ell}+{\ell\over 2}P_1^TP_1^R\Bigr)+x_1^R\Bigl[{1-e^{-\ell P_0^T}\cosh \ell P_0^R \over \ell}+{\ell \over 4}(P_1^{T^2}+P_1^{R^2})\Bigr]-x_0^TP_1^R-x_0^RP_1^T.
\end{split}
\end{align}
Since we can easily go from the set of generators $\{P_\mu^{A,B},N^{A,B}\}$ to the set $\{P_\mu^{T,R},N^{T,R}\}$ and viceversa, these show that the set $\{P_\mu^{T,R},N^{T,R}\}$ forms a closed algebra of symmetry. For the
commutators between the boost and the time translation generators one straightforwardly finds
\begin{align}
\begin{split}
&[N^T,P_0^T]=[N^A+N^B,P_0^A+P_0^B]=iP_1^T,\\
&[N^T,P_0^R]=[N^A+N^B,P_0^A-P_0^B]=iP_1^R,\\
&[N^R,P_0^T]=[N^A-N^B,P_0^A+P_0^B]=iP_1^R,\\
&[N^R,P_0^R]=[N^A-N^B,P_0^A-P_0^B]=iP_1^T \, .
\end{split}
\end{align}
The derivations of the remaining commutators are a little more involved but still manageable. For instance
one has:
\begin{align}
\begin{split}
[N^T,P_1^T]&=[N^A,P_1^A]+[N^B,P_1^B]\\
&={1-e^{-2\ell P_0^A} \over 2\ell}-{\ell \over 2}P_1^{A^2}+{1-e^{-2\ell P_0^B} \over 2\ell}-{\ell \over 2}P_1^{B^2}\\
&={1-{e^{-2\ell P_0^A}+e^{-2\ell P_0^B} \over 2}\over \ell}-{\ell \over 2}(P_1^{A^2}+P_1^{B^2})\\
&={1-e^{-\ell P_0^T}\cosh \ell P_0^R \over \ell}-{\ell \over 4}(P_1^{T^2}+P_1^{R^2}) \, .
\end{split}
\end{align}
and in similar fashion:
\begin{align}
\begin{split}
[N^T,P_1^R]&=[N^A,P_1^A]-[N^B,P_1^B]\\&={1-e^{-2\ell P_0^A} \over 2\ell}-{\ell \over 2}{P_1^A}^2-{1-e^{-2\ell P_0^B} \over 2\ell}+{\ell \over 2}{P_1^B}^2\\&
={-e^{-2\ell P_0^A}+e^{-2\ell P_0^B} \over 2\ell}-{\ell \over 2}({P_1^A}^2-{P_1^B}^2)\\&=e^{-\ell P_0^T}{\sinh \ell P_0^R \over \ell}-{\ell \over 2}P_1^T P_1^R.
\end{split}
\end{align}
It is easy to realize that the remaining two commutators are actually identical to the previous ones, since they end up connecting the single particle generators $A$ or $B$ with the same signs $+$ or $-$, that is
\begin{equation}[N^R,P_1^T]=[N^T,P_1^R], \qquad \qquad
[N^R,P_1^R]=[N^T,P_1^T].
\end{equation}
Finally it is not hard to compute the coproducts of the total and relative boost generators:
\begin{align}
\begin{split}
&\Delta(N^T)=N^T\otimes 1+e^{-{\ell\over 2}P_0^T}\cosh{\ell \over 2}P_0^R\otimes N^T-e^{-{\ell\over 2}P_0^T}\sinh{\ell \over 2}P_0^R\otimes N^R\\
&\Delta(N^R)=N^R\otimes 1+e^{-{\ell\over 2}P_0^T}\cosh{\ell \over 2}P_0^R\otimes N^R-e^{-{\ell\over 2}P_0^T}\sinh{\ell \over 2}P_0^R\otimes N^T,
\end{split}
\end{align}
which unsurprisingly have the same form of the coproducts \eqref{coprodottilinear}.
Notice again that for small relative momentum, both the commutators and the coproducts tend to the $2\kappa$-Poincaré ones.

We therefore have a description of the relativistic
symmetries of our two-particle system in terms of center-of-mass degrees of freedom
and degrees of freedom of the motion relative to the center of mass which is given
by a robust Hopf algebra, with generators $\{P_0^T,P_0^R,P_1^T,P_1^R,N^T,N^R\}$. In cases where there is negligible
motion relative to the center of mass the relativistic symmetries are fully described by
 a $2\kappa$-Poincaré Hopf algebra, with the prefix ''$2\kappa$"
signaling the fact that, denoting with $\ell$ the deformation parameter of the $\kappa$-Poincaré Hopf algebra
governing the single-particle case, for the center of mass of a two-particle system we have exactly the same symmetries
(if indeed there is negligible
motion relative to the center of mass)
but with halved deformation parameter, ${\ell \over 2}$.

In closing this section we want to provide further (and more explicit) observations
concerning the emergence of the $2\kappa$-Poincaré Hopf algebra.
We start by noticing that, in light of results established above, as basis for the space of fields describing
two identical particles in $\kappa$-Minkowski one can equivalently adopt either
\begin{equation}
e^{ik_1^A x_1^A}e^{-ik_0^A x_0^A}e^{ik_1^B x_1^B}e^{-ik_0^B x_0^B}.
\end{equation}
or
\begin{equation}
e^{ik_1^T x_1^T}e^{ik_1^R x_1^R}e^{-ik_0^T x_0^T}e^{-ik_0^R x_0^R} \, .
\end{equation}
Then a generic function describing the
two particles can be written as
\begin{equation}
g(x_1^T, x_0^T, x_1^R, x_0^R) = \int d^2 k^T d^2 k^R \tilde{g}(k_1^T, k_0^T, k_1^R, k_0^R)  e^{ik_1^T x_1^T}e^{ik_1^R x_1^R}e^{-ik_0^T x_0^T}e^{-ik_0^R x_0^R} \, .
\end{equation}
The subspace $C^T$ of fields which depend exclusively on the center-of-mass coordinates can be characterized by restricting  the analysis to functions $\tilde{g}$ such that $\tilde{g}(k_1^T, k_0^T, k_1^R, k_0^R)=0$ whenever $k_1^R \neq 0$
and/or $k_0^R \neq 0$. On this subspace of fields the relativistic symmetries are fully described
by the $2k$-Poincaré Hopf algebra, and in particular for such fields the boost generator
admits the following simple description
\begin{equation}
N^T=x_1^T\Bigl({1-e^{-\ell P_0^T} \over \ell}+{\ell \over 4}P_1^{T^2}\Bigr)-x_0^T P_1^T \, .
\end{equation}

\section{Summary and outlook}
\label{sec5}
The study we here reported intends to contribute to the search of the proper description of multi-particle
systems in some quantum spacetimes, using the $\kappa$-Minkowski spacetime as reference example.
We obtained new results for both of the most studied definitions of total momentum
in $\kappa$-Minkowski. In particular, we observed that the definition of total momentum which is inspired
by the $\kappa$-Poincaré coproduct is even more concerning than previously appreciated:
we found that besides being, as already known, based on nonlinearities that could be pathological for macroscopic bodies,
it also leads to a paradoxical description of center-of-mass coordinates and/or a picture in which there is no
closed (deformed) Heisenberg algebra for the center-of-mass motion.
The results we obtained for the alternative undeformed definition of total momentum appear
to be more encouraging: not only the total momentum does not involve nonlinearities but also the full
description of relativistic symmetries appears to be well behaved as the number of particles
in the system grows (signaled in our study by the emergence of a $2\kappa$-Poincarè description
for the two-particle system from the $\kappa$-Poincarè description of the composing particles);
moreover, one gets a closed (deformed) Heisenberg algebra for the center-of-mass motion.
The only residual peculiarity present when adopting the undeformed definition of total momentum
resides in an incomplete separation between center-of-mass motion and motion relative to the center of mass,
which however does not appear to be alarming since its form would be inconsequential in typical many-particle systems,
whose relative motion
is strongly suppressed with respect to the center-of-mass motion.
Nonetheless,  these limitations to the separation between center-of-mass motion and motion relative to the center of mass
are noteworthy from a conceptual perspective, and we conjecture that they will be found also in other quantum
spacetimes. We feel that some priority should be given to
the development of suitable techniques of analysis of this issue adapted to
the different quantum spacetimes that might be affected by it.

Evidently our findings also provide motivation for further investigation of another long-standing issue
for these scenarios with deformed relativistic symmetries, which concerns the possibility of nonuniversal relativistic behavior:
if a two-particle composite system is governed by $2\kappa$-Poincarè relativistic properties while single particles are
governed by $\kappa$-Poincarè relativistic properties then the description of processes involving both
a two-particle composite system and a fundamental particle should require a sophisticated picture that takes into account
the different relativistic properties. Preliminary results suggest that
this can be done satisfactorily (see, {\it e.g.}, Refs.~\cite{nonuniversal1, nonuniversal2}),
but more in-depth analyses are needed.

\section*{Acknowledgements}

G.A.-C.'s work on this project was supported by the FQXi grant 2018-190483 and by the MIUR, PRIN 2017 grant 20179ZF5KS.

\end{flushleft}


\begin{thebibliography}{80}
\addcontentsline{toc}{chapter}{Bibliografia}
\bibitem{amelino2002IJMPD} G. Amelino-Camelia, \textit{Relativity in space-times with short-distance structure governed by an observer-independent (Planckian) length scale}, International Journal of Modern Physics D, Vol. 11, N. 01, pp. 35-59 (2002)
\bibitem{DFR1995} S. Doplicher, K. Fredenhagen, J. E. Roberts, \textit{The quantum structure of spacetime at the Planck scale and quantum fields},Communications in Mathematical Physics 172.1: 187-220 (1995)
\bibitem{BergSmit1982} P.G. Bergmann, G.J. Smith, \textit{Measurability analysis of the linearized gravitational
field}, Gen. Rel. Grav. 14 1131-1166 (1982)
\bibitem{MajRue1994} S. Majid, H. Ruegg, \textit{Bicrossproduct structure of $\kappa$-Poincaré group and non-commutative geometry}, Physics Letters B 334.3-4: 348-354 (1994)
\bibitem{HeisAlg} J. Lukierski, H. Ruegg, W.J. Zakrzewski. \textit{Classical and quantum mechanics of free $\kappa$-relativistic systems}, Annals of Physics 243.1: 90-116 (1995)
\bibitem{GGV2013} G. Amelino-Camelia, V. Astuti, G. Rosati, \textit{Relative locality in a quantum spacetime and the pregeometry of $\kappa$-Minkowski}, The European Physical Journal C 73.8: 2521 (2013)
\bibitem{reloc2011} G. Amelino-Camelia, L. Freidel, J. Kowalski-Glikman, L. Smolin, \textit{Principle of relative locality}, Physical Review D 84.8: 084010 (2011)
\bibitem{reloc2013} G. Gubitosi, F. Mercati, \textit{Relative locality in $\kappa$-Poincaré}, Classical and Quantum Gravity 30.14: 145002 (2013)
\bibitem{freidelElivinePRL} L. Freidel, E. R. Livine, \textit{3D quantum gravity and effective noncommutative quantum field theory}, Physical review letters 96.22: 221301 (2006)
\bibitem{gacLivingReview} G. Amelino-Camelia, \textit{Quantum spacetime phenomenology}, Living Rev. Rel. 16 5 (2013)
\bibitem{Soccer 1} G. Amelino-Camelia, L. Freidel, J. Kowalski-Glikman, L. Smolin, \textit{Relative locality and the soccer ball problem}, Phys.Rev. D84 087702 (2011)
\bibitem{Soccer 2} S. Hossenfelder, \textit{The Soccer-Ball Problem}, SIGMA 10 (2014)
\bibitem{Soccer 3} G. Amelino-Camelia, \textit{Planck-Scale Soccer-Ball Problem: A Case of Mistaken Identity}, Entropy 19 no.8, 400 (2017)
\bibitem{nonlin1} T. Jacobson, S. Liberati, D. Mattingly, \textit{Lorentz violation at high energy: Concepts, phenomena and astrophysical constraints}, Ann. Phys. 321, 150–196 (2006)
\bibitem{nonuniversal1} G. Amelino-Camelia, \textit{Particle-Dependent Deformations of Lorentz Symmetry}, Symmetry, 4(3), 344-378 (2012)
\bibitem{nonuniversal2} M. Palmisano, G. Amelino-Camelia, M. Ronco, G. D'Amico, \textit{Mixing coproducts for theories with particle-dependent relativistic properties}, International Journal of Modern Physics D, Vol. 29, Issue N. 02 2050017 (2020)
\end{thebibliography}
\end{document}